\documentclass[12pt]{article}
\usepackage{epsfig}
\begin{document}
\title{\bf Distances in plane membranes \\ } 
\author{ A.F.F. Teixeira \thanks{teixeira@cbpf.br} \\
         {\small Centro Brasileiro de Pesquisas F\'{\i}sicas } \\
         {\small 22290-180 Rio de Janeiro-RJ, Brazil} \\ 
       }
\maketitle

\def\Pl{${\cal P}(l)$ }  \def\bea{\begin{eqnarray}}  \def\eea{\end{eqnarray}}

\begin{abstract}
A flat membrane with given shape is displayed;  
two points in the membrane are randomly selected;  
the probability that the separation between the points have a specific value is sought. 
A simple method to evaluate the probability density is developed, 
and is easily extended to spaces with more dimensions.  
\end{abstract} 

\section{Introduction}
A most common geometrical problem encountered in exact and natural sciences 
(engineering, physics, chemistry, biology, etc) is: 
given a surface with finite area and definite form, 
and randomly choosing two points in it, find the probability density \Pl   
that these points have a prescribed separation $l$. 

To see the relevance of the subject consider the following example taken from biology: 
a live membrane is infected at some spots, and the progress of the infection  
is suspected to depend on the mutual separation between the infected points. 
A knowledge of the mean separation $l_{mean}$, the mean inverse separation 
$(l^{-1})_{mean}$, and the mean squared inverse separation $(l^{-2})_{mean}$ between 
points in the membrane,
all depend on the probability density, and are crucial to an analysis of the process. 

In this note we find the functions \Pl for three finite surfaces widely dealt with: 
the circle, the square, and the rectangle. 
{}From these examples the method for obtaining \Pl for other surfaces is trivially inferred. 
The method can also be easily extended to three-dimensional spaces, such as spherical drops, 
or parallelepipeds. 
  
\section{Circular membranes} 
In a plane disk with diameter $\delta$ two points $A$, $B$ are randomly chosen. We want the probability \Pl$dl$ that the separation between the points lies between $l$ and $l+dl$. 
The probability density \Pl has to satisfy the normalization condition 
\bea                                                        \label{1}
\int_0^\delta{{\cal P}(l)dl}=1 .
\eea 

Our first concern is: having chosen a point $A$ in the ({\it infinite}) euclidian plane, 
next randomly choosing another point $B$ also in the plane, 
we seek the probability that their separation have a value $l$. 
That probability clearly is proportional to the measure of the {\it locus} of $B$ 
-- the circle with center $A$ and radius $l$ -- so the locus has measure $2\pi l$. 
However, since neither the finiteness of the disk nor its shape were taken into account, 
the analysis of the problem is still incomplete.   

We have to examine all possible line segments entirely embedded in the disk. 
As a matter of fact, the symmetry of the disk permits restrict the study to segments 
aligned in just one direction; we choose the vertical direction, for definiteness. 

\vspace*{3mm} 
\centerline{ \epsfig{file=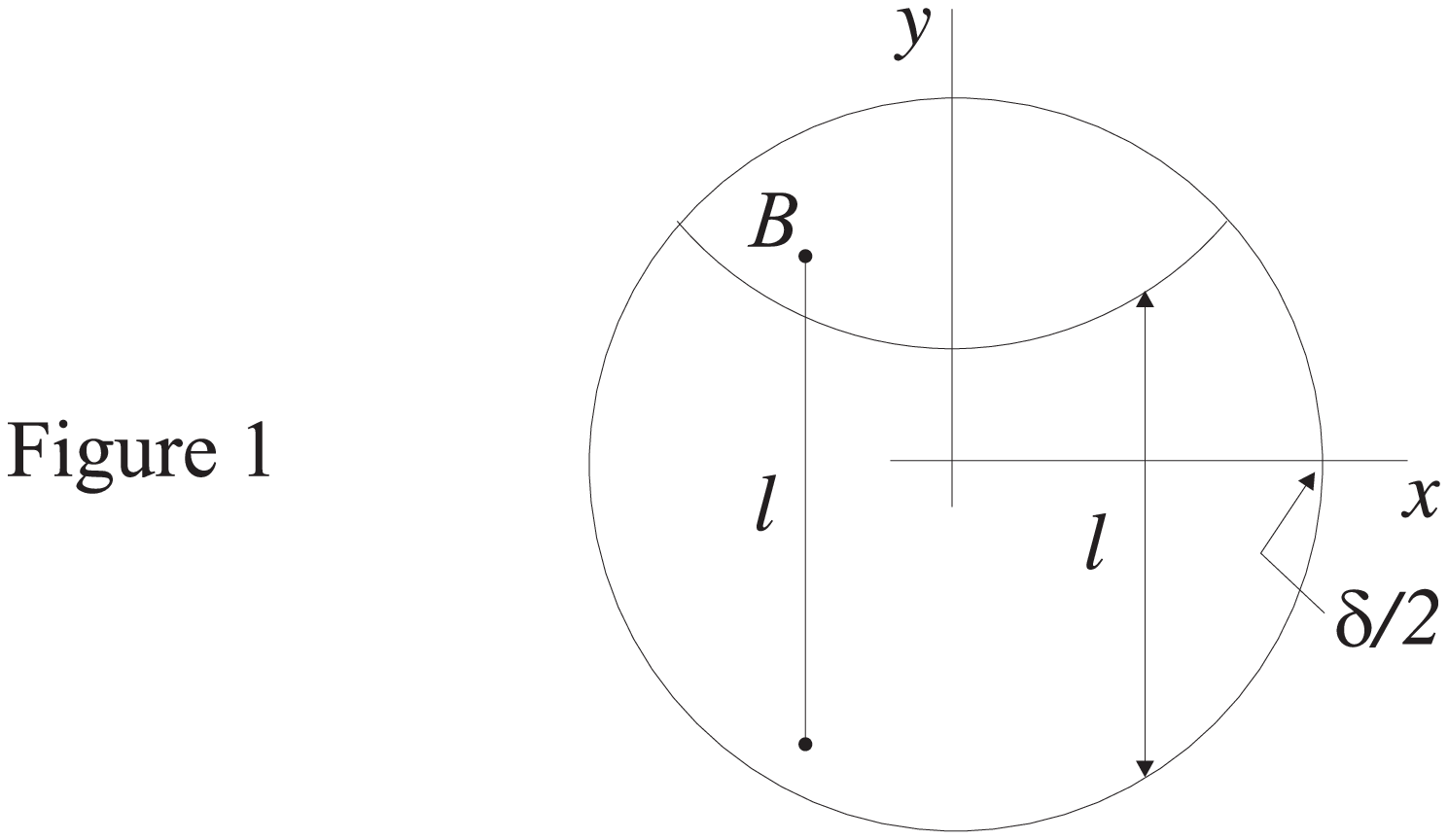, width=7cm, height=4cm} }
\vspace*{3mm} 

In figure 1 we note that the upper tip $B$ of a vertical segment with length $l$ 
has to lie in the lens shaped region enclosed by the circles $x^2+y^2=(\delta/2)^2$ 
and $x^2+(y-l)^2=(\delta/2)^2$. 
The area of this lens is 
\bea                                                         \label{2}
S(l)=\frac{1}{2}\left[\delta^2\cos^{-1}(l/\delta) - l\sqrt{\delta^2-l^2}\right] .
\eea
Writing the probability density as 
${\cal P}(l)=k\,lS(l)$ and imposing the normalization condition (\ref{1}) we find 
$k=32/(\pi\delta^4)$, and finally 
\bea                                                          \label{3}
{\cal P}(l)=\frac{16\,l}{\pi\delta^4}\left[\delta^2\cos^{-1}(l/\delta) - l\sqrt{\delta^2-l^2}\right] .
\eea
A graph of \Pl is given in the figure 2; also a normalized histogram obtained via computer simulation is reproduced, to give confidence in the calculations.  

\vspace*{3mm} 
\centerline{ \epsfig{file=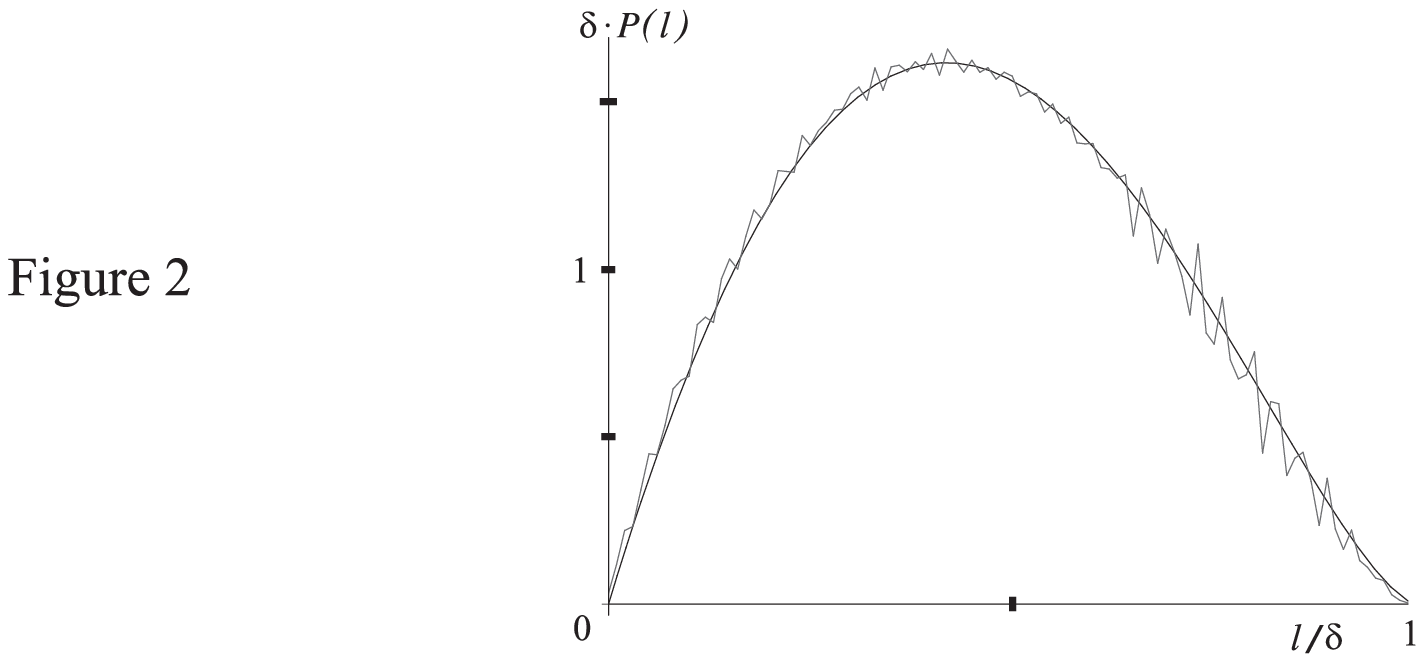, width=10cm, height=5cm} }
\vspace*{3mm} 

\section{Square membranes}
Similarly as before, we randomly choose two points in a square with side $a$, and want 
the probability density that their separation be $l$; the normalization condition now reads 
\bea                                                            \label{4}
\int_0^{a\sqrt{2}}{{\cal P}(l)dl}=1 .
\eea 
Again an overall multiplicative factor $l$ is expected in the expression of \Pl, 
and we are next interested in the line segments lying entirely inside the square. 

We initially consider the segments with length $l\!<\!a$; in this case  
the symmetries of the square allow reduce our study to the segments with slope 
lying between $\phi=0$ and $\phi=\pi/4$, as is seen in the first figure 3. 

\vspace*{3mm} 
\centerline{ \epsfig{file=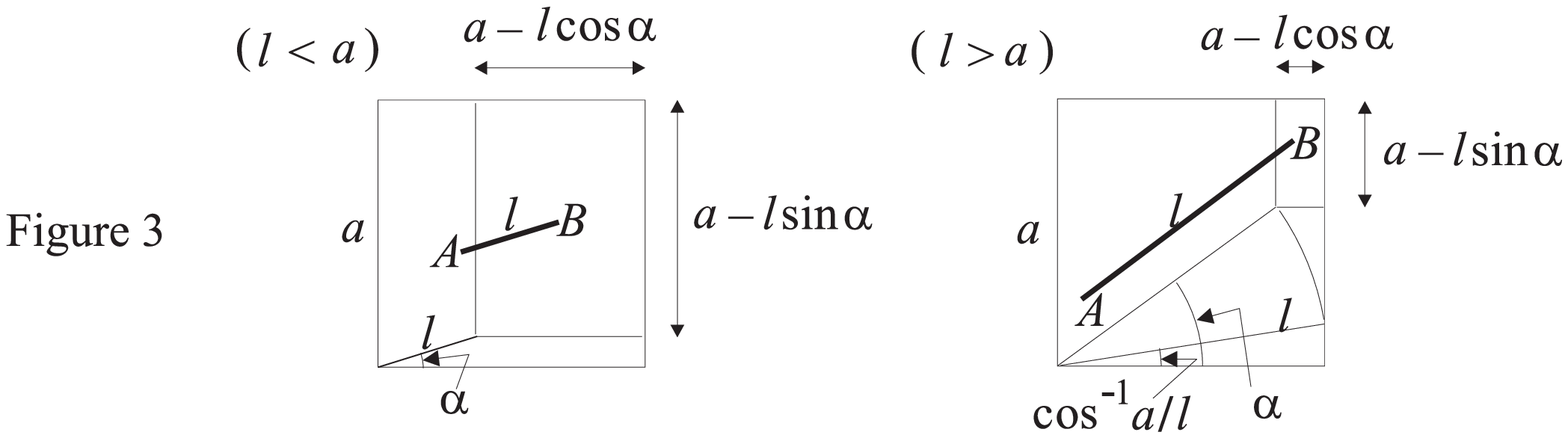, width=15cm, height=45mm} }
\vspace*{3mm} 

\noindent The locus of the upper tip $B$ of a segment with slope $\phi$ and length 
$l$ is a rectangle with area 
\bea  										\label{5}
S(l,\phi)=(a-l\cos\phi)(a-l\sin\phi). 
\eea
Taking into account all slopes in the range $[0, \pi/4]$ we write 
\bea 											\label{6}		
{\cal P}(l\!<\!a)=k\,l\int_0^{\pi/4}{S(l,\phi)d\phi} . 
\eea

For line segments with length $l\!>\!a$ the minimum slope is $\cos^{-1}a/l$, as is clear 
in the second figure 3, so now 
\bea 											\label{7}
{\cal P}(l\!>\!a)=k\,l\int_{\cos^{-1}a/l}^{\pi/4}{S(l,\phi)d\phi} . 
\eea
The normalization constant is obtained from (\ref{4}), 
\bea    										\label{8}		
\int_0^a{{\cal P}(l\!<\!a)dl} + \int_a^{a\sqrt{2}}{{\cal P}(l\!>\!a)dl}=1, 
\eea 
and has value $k=8/a^4$. We then have \Pl given by the two expressions 
\bea                                                              \label{9}
{\cal P}(l\!<\!a)=\frac{2l}{a^4} \left[ l^2-4al+\pi a^2\right] , 
\eea 
\bea                                                              \label{10}
{\cal P}(l\!>\!a)=\frac{2l}{a^4} \left[ 4a\sqrt{l^2-a^2}+4a^2\sin^{-1}a/l-l^2-\pi a^2-2a^2\right] . 
\eea
A graph of \Pl is given in the figure 4; also a normalized histogram obtained via computer simulation is superimposed for comparison. 

\vspace*{3mm} 
\centerline{ \epsfig{file=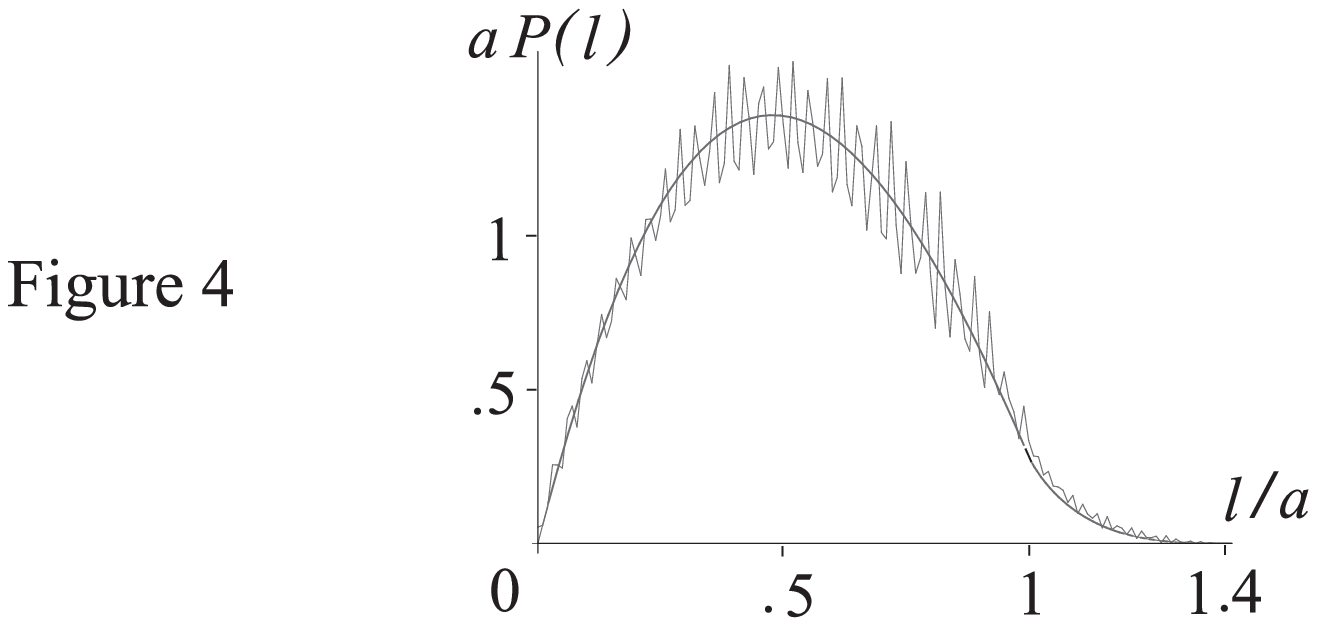, width=10cm, height=5cm} }
\vspace*{3mm} 

\section{Rectangular membranes}
We assume a rectangle with sides $a$ and $b\!<\!a$; to investigate separations $l$ 
between points in the rectangle we need now distinguish three different possibilities, 
depending on the value of $l$ relative to $a$ and $b$. See figure 5. 

\vspace*{3mm} 
\centerline{ \epsfig{file=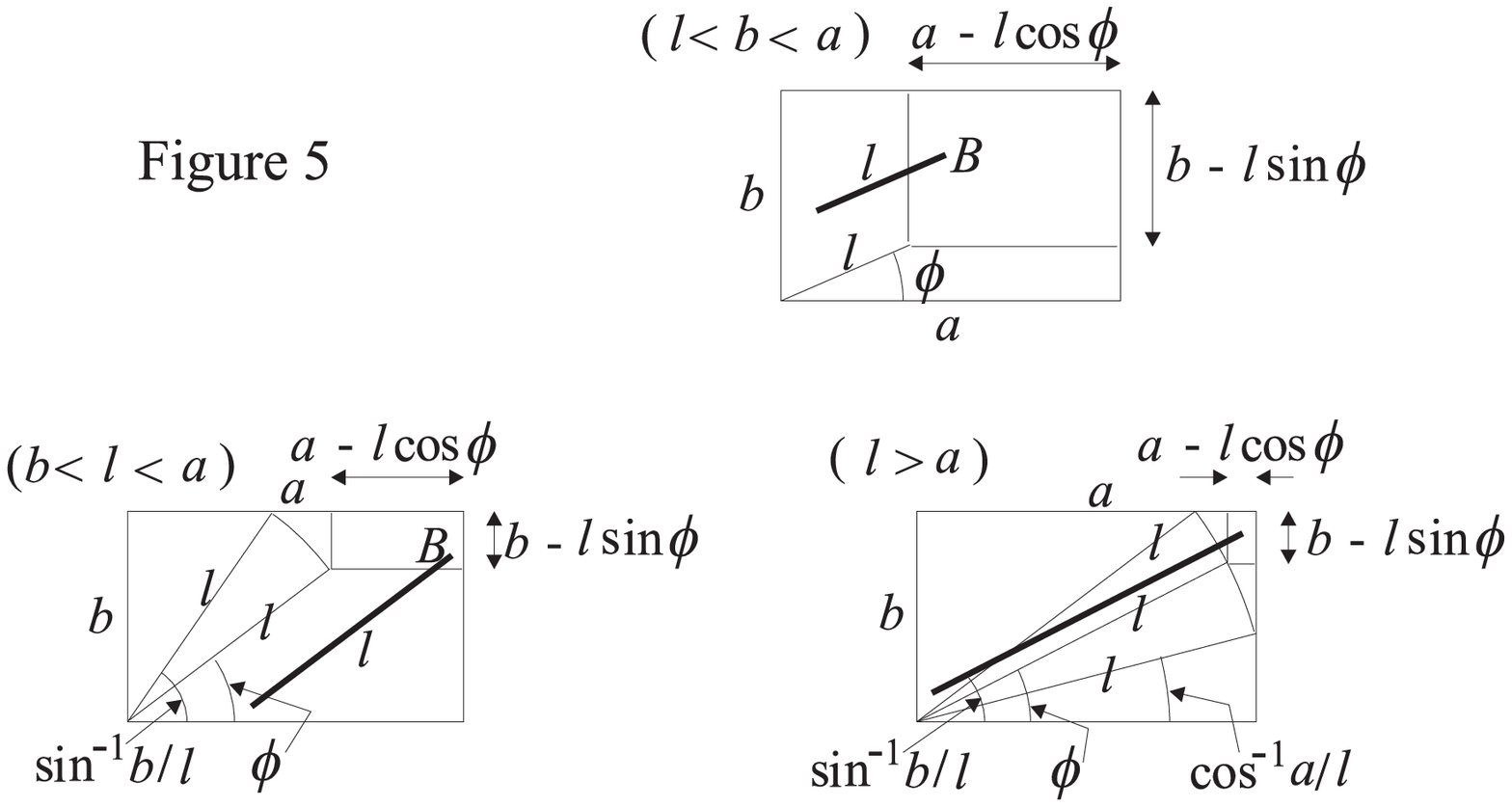, width=13cm, height=7cm} }
\vspace*{3mm} 

In all cases the angular probability density ${\cal P}(l,\phi)$ 
is similar as before,  
\bea  										\label{11}
{\cal P}(l,\phi)=k\,l(a-l\cos\phi)(b-l\sin\phi). 
\eea 
The symmetries of the rectangle permit restrict the study to segments with slope 
from $\phi=0$ to $\phi=\pi/2$. 
When $l\!<\!b\!<\!a$ we integrate (\ref{11}) from $\phi=0$ to $\phi=\pi/2$ 
and obtain ${\cal P}(l\!<\!b)$.  
When $b\!<\!l\!<\!a$ the maximum slope is reduced to $\phi=\sin^{-1}b/l$, and 
the integration gives ${\cal P}(b\!<\!l\!<\!a)$. 
Finally, when $l\!>\!a\!>\!b$ the slope ranges from $\phi=\cos^{-1}\,a/l$ to 
$\phi=\sin^{-1}\,b/l$, and the integration gives ${\cal P}(l\!>\!a)$. 
 
The normalization constant $k$ is still unassigned; to fix it we impose 
the normalization condition 
\bea                                                               \label{12}
\int_0^b{{\cal P}(l\!<\!b)dl} + \int_b^a{{\cal P}(b\!<\!l\!<\!a)dl} + \int_a^{\sqrt{a^2+b^2}}{{\cal P}(l\!>\!a)dl}=1 ,
\eea
and find $k=4/(ab)^2$; the probability density \Pl is then expressed in the three stages 
\bea                                                               \label{13} 
{\cal P}(l\!<\!b)=\frac{4\,l}{a^2b^2}\left[l^2/2-(a+b)l+\pi ab/2\right], 
\eea 
\bea                                                               \label{14} 
{\cal P}(b\!<\!l\!<\!a)=\frac{4\,l}{a^2b^2}\left[ab\sin^{-1}b/l-al+a\sqrt{l^2-b^2}-b^2/2\right], 
\eea
\bea                                                               \nonumber                                                         
{\cal P}(l\!>\!a)=\frac{4\,l}{a^2b^2}\hskip2mm
[ab(\sin^{-1}a/l+\sin^{-1}b/l)+ 					        
\eea
\bea                                                                \label{15}
a\sqrt{l^2-b^2}+b\sqrt{l^2-a^2}-l^2/2-\pi ab/2-(a^2+b^2)/2\hskip1mm]. 
\eea
A graph of \Pl is given in the figure 6, drawn for $b=a/2$; also a normalized histogram 
obtained via computer simulation is given as illustration. 

\vskip3mm
\centerline{ \epsfig{file=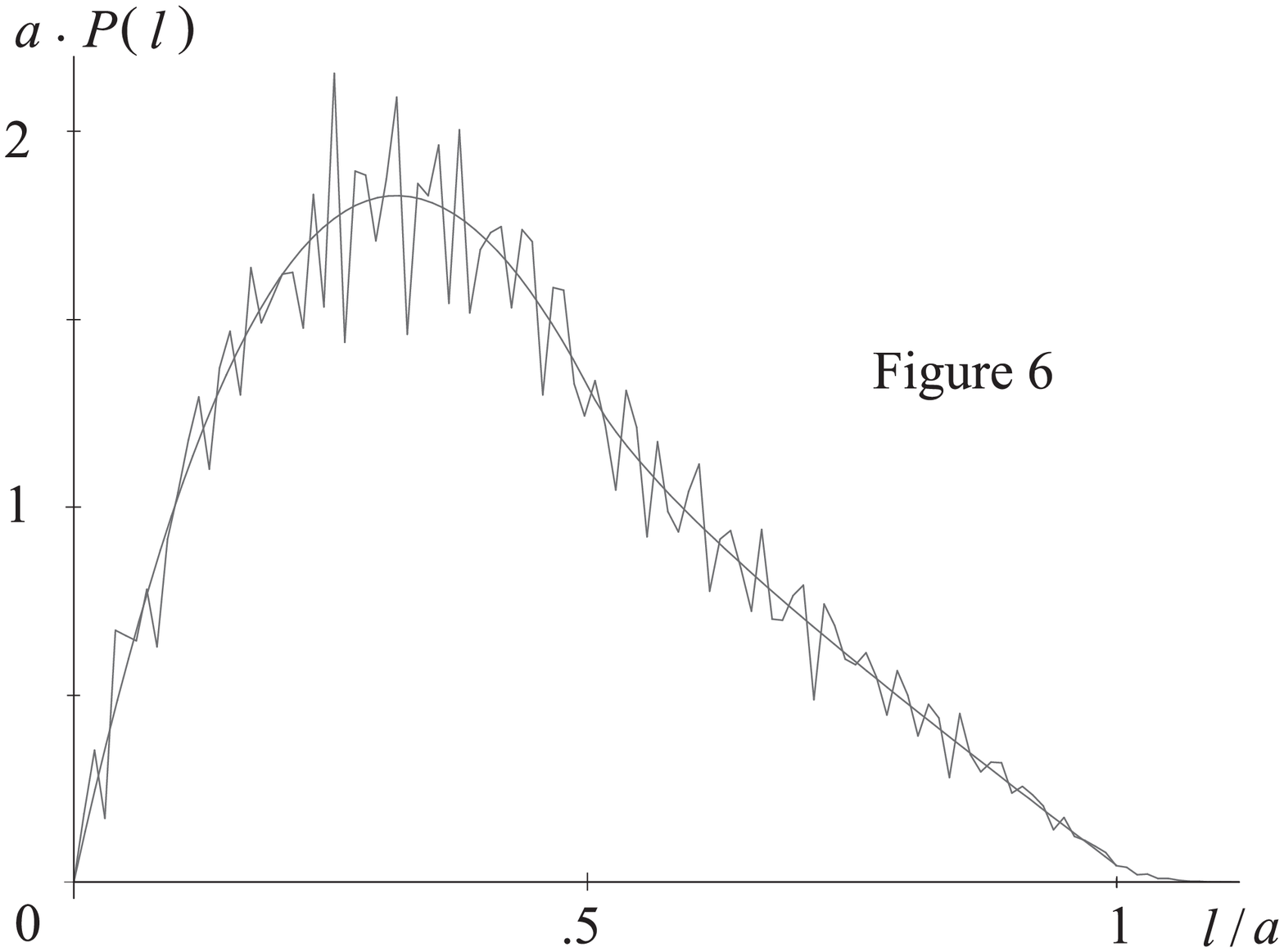, width=9cm, height=65mm} }
\vskip3mm

\section{Discussion}
The problem of investigating separations between points in a given space \cite{Beres} 
recently received a strong and unexpected stimulus coming from cosmology 
\cite{FarrarMelott}; more specifically, from the so-called cosmic crystallography, 
which aims to unveil the shape of the universe \cite{LeLaLu} -- \cite{Janna}.

The algorithm formerly used in the present note to find the various probability 
densities \Pl closely followed that of cosmologists. 
However, it soon became evident that a new approach was imperative to circumvent the 
long calculations arising from that algorithm, when applied to spaces different from balls. 

The idea of using normalized probability densities \Pl greatly simplifies the task 
of obtaining mean quantities concerning separations; {\it e.g.}, we have 
\bea                                                          \label{16} 
l_{mean}=\int_0^{l_{max}}{l\,{\cal P}(l)\,dl}, \hskip3mm 
(l^{-1})_{mean}=\int_0^{l_{max}}{l^{-1}\,{\cal P}(l)\,dl} ; 
\eea
these are particular instances of the general rule  
\bea                                                          \label{17} 
[f(l)]_{mean}=\int_0^{l_{max}}{f(l)\,{\cal P}(l)\,dl}.
\eea 

In figures 2, 4, and 6 both abscissa and ordinate were chosen dimensionless; 
by simple inspection we then confirm that the corresponding funtions \Pl are indeed 
normalized. 

In figure 2 we note that the most probable separation in the disk with diameter $\delta$ 
is $l_{mp}\approx 0.42\delta$, corresponding to the value of $l$ where the function \Pl 
of (\ref{3}) is maximum; while the mean separation is slightly greater, 
$l_{mean}\approx 0.45\delta$. We also note that the function changes curvature at 
$l/\delta=\sqrt{2/3}\approx 0.82$. 

In figure 4 we see that the most probable separation between points in a square 
with side $a$ is $l_{mp}\approx 0.48a$, while the mean separation is 
$l_{mean}\approx 0.52a$. 
Although both the function \Pl in (\ref{9}) and (\ref{10}) and its first derivative 
are continuous at $l=a$, the second derivative is not: in fact, 
$a^2d^2{\cal P}(l)/dl^2$ abruply changes from the finite negative value $-4$ when $l=a-\epsilon/2$ 
to the diverging positive value $8/\sqrt{\epsilon}$ when $l=a+\epsilon/2$. 
Of course the curvature of \Pl changes sign at $l=a$. 
We still note in figure 4 the rapidly decreasing density of separations when $l/a$ 
approaches $\sqrt{2}$; this was already expected, since these large separations 
correspond to segments with both endpoints in diagonally opposite corners of the square, 
and corner regions are small in comparison with the whole square. 

In figure 6, corresponding to the rectangle, we again note that both \Pl of (\ref{13})-(\ref{15}) and its first derivative are continuous throughout $0\!<\!l\!<\!\sqrt{a^2+b^2}$. 
And again the second derivative shows infinite discontinuity, now at $l\!=\!b$ and 
also at $l\!=\!a$; nevertheless \Pl changes curvature only at $l\!=\!b$. 
When $b$ diminishes relative to $a$ we find that the graph of \Pl gradually resembles 
a right triangle; when $b/a\to 0$ the plot is a straight line going from (0, 2) to 
(1, 0), as in the figure 6 of ref. \cite{Beres} or figure 3 of ref. \cite{T1}: namely, 
$a\,{\cal P}(l)=2(1-l/a)$. 

In extending the present note to three-dimensional euclidean spaces one should replace 
the overall multiplicative factor $l$ in the probability density with a factor $l^2$; 
this is because the locus of the points that are at a distance $l$ from a fixed point 
in three-space is a two-dimensional sphere, whose area $4\pi l^2$ increases with $l^2$. 

\section{Acknowledgments}
Are due to Germ\'{a}n I. Gomero and Marcelo J. Rebou\c{c}as for pointing out various 
important references and for fruitful conversations.


\begin{thebibliography}{20}

\bibitem{Beres} Krzysztof Bere\'{s}, ``Distance distribution'', {\cal Zeszyty Naukowe 
        Universytetu Jagiello\'{n}skiego - Acta Cosmologica - Z. 5} (1976) 7-27 
\bibitem{FarrarMelott} Kelly A. Farrar and Adrian L. Melott, ``Gravity in twisted space'', 
        {\cal Computers in Physics}, Mar/Apr 1990, 185-189
\bibitem{LeLaLu} Roland Lehoucq, M. Lachi\`{e}ze-Rey and Jean-Pierre Luminet, ``Cosmic 
        crystallography'', gr-qc/9604050 
\bibitem{FagGaus1} Helio V. Fagundes and Evelise Gausmann, ``On closed Einstein-de Sitter 
        universes'', astro-ph/9704259 
\bibitem{LeLuUz} Roland Lehoucq, Jean-Pierre Luminet and Jean-Philippe Uzan, ``Topological 
        lens effects in universes with non-euclidian compact spatial sections'', 
        astro-ph/9811107 
\bibitem{GTRB1} Germ\'{a}n I. Gomero, Antonio F.F. Teixeira, Marcelo J. Rebou\c{c}as 
        and Armando  Bernui, ``Spikes in cosmic crystallography'', gr-qc/9811038 
\bibitem{FagGaus2} Helio V. Fagundes and Evelise Gausmann, ``Cosmic crystallography in 
        compact hyperbolic universes'', astro-ph/9811368 
\bibitem{LuRouk} Jean-Pierre Luminet and Boudewijn F. Roukema, ``Topology of the universe: 
         theory and observation'', astro-ph/9901364 
\bibitem{UzLeLu} Jean-Philippe Uzan, Roland Lehoucq and Jean-Pierre Luminet ``A new 
         crystallographic method for detecting space topology'', astro-ph/9903155
\bibitem{BT1} Armando Bernui and Antonio F.F. Teixeira ``Cosmic crystallography: 
        three multipurpose functions'', astro-ph/9904180
\bibitem{GRT1} Germ\'{a}n I. Gomero, Marcelo J. Rebou\c{c}as and Antonio F.F. Teixeira, 
       ``Spikes in cosmic crystallography II: topological signature of compact flat 
         universes'', gr-qc/9909078
\bibitem{GRT2} Germ\'{a}n I. Gomero, Marcelo J. Rebou\c{c}as and Antonio F.F. Teixeira, 
        ``A topological signature in cosmic topology'', gr-qc/9911049
\bibitem{BT2} Armando Bernui and Antonio F.F. Teixeira ``Cosmic crystallography: 
         the euclidean isometries'', gr-qc/0003063 
\bibitem{T1} Antonio F.F. Teixeira ``Cosmic crystallography in a circle'', gr-qc/0005052  
\bibitem{LeUzLu2} Roland Lehoucq, Jean-Pierre Luminet and Jean-Philippe Uzan, 
        ``Limits of crystallographic methods for detecting space topology'', astro-ph/0005515
\bibitem{T2} Antonio F.F. Teixeira ``Cosmic crystallography: the hyperbolic isometries'',
        gr-qc/0010107 
\bibitem{GRT3} Germ\'{a}n I. Gomero, Marcelo J. Rebou\c{c}as and Antonio F.F. Teixeira,  
        ``Signature for the shape of the universe'', gr-qc/0105048
\bibitem{EvLeLuUzWe} Evelise Gausmann, Roland Lehoucq, Jean-Pierre Luminet, Jean-Philippe 
         Uzan and Jeffrey Weeks, ``Topological lensing in spherical spaces'', gr-qc/0106033
\bibitem{Janna} Janna Levin ``Topology and the cosmic microwave background'', gr-qc/0108043

 
 



\end{thebibliography}
\end{document}